\documentclass[twocolumn,showpacs,preprintnumbers,amsmath,amssymb,superscriptaddress]{revtex4}
\usepackage{graphicx}
\usepackage{xcolor}
\usepackage{cases}
\DeclareMathOperator\erf{erf}
\DeclareMathOperator\erfc{erfc}

\begin{document}
\title{
Random walks with fractally correlated traps:\\ Stretched exponential and power law survival kinetics  
}
\author{Dan Plyukhin}
\email{dplyukhin@cs.toronto.edu} \affiliation{ Department of
  Computer Science, University of Toronto, Toronto, ON, Canada}
\author{Alex V. Plyukhin}
\email{aplyukhin@anselm.edu} \affiliation{
  Department of Mathematics, Saint Anselm College, Manchester, NH,
  USA}
\date{\today}

\begin{abstract}

We consider the survival probability $f(t)$ of a random walk with a constant
hopping rate $w$ on a host lattice of fractal dimension $d$ and spectral
dimension $d_s\le 2$, with spatially correlated traps. The traps form a
sublattice with fractal dimension $d_a<d$ and are characterized by the
absorption rate $w_a$ which may be finite (imperfect traps) or infinite
(perfect traps). Initial coordinates are chosen randomly at or within a fixed
distance of a trap. For weakly absorbing traps ($w_a\ll w$), we find that $f(t)$
can be closely approximated by a stretched exponential function over the initial
stage of relaxation, with stretching exponent $\alpha=1-(d-d_a)/d_w$, where
$d_w$ is the random walk dimension of the host lattice.
At the end of this initial stage there occurs a crossover to power law kinetics
$f(t)\sim t^{-\alpha}$ with the same exponent $\alpha$ as for the stretched exponential regime. For strong absorption $w_a\agt w$, including the limit of
perfect traps $w_a\to \infty$, the stretched exponential regime is absent 
and the decay of $f(t)$ follows, after a short transient,  
the aforementioned power law for all times.

\end{abstract}


\maketitle

\section{Introduction}

The paradigm of random walks with correlated traps is relevant to many interdisciplinary problems (random search strategies, exciton trapping by polymer chains, ligand binding to receptors on cell surfaces, foraging patterns, 
etc.), but it is also interesting from the more general perspective of the dynamics of complex systems with correlated disorder ~\cite{Stillinger}. Many previous works concerned trapping kinetics for regularly distributed traps~\cite{Kenkre}, some specific types of finite-range correlations and clusters~\cite{clusters,clusters1,clusters2,Agliari}, as well as traps distributed with critical (long-ranged) positional correlations~\cite{Nakanishi}. 

In this paper we consider random walks on a host lattice with spectral dimension $d_s\le 2$ (where diffusion is compact~\cite{deGennes,Rammal}) with correlated imperfect traps forming a proper fractal sublattice. For such a model, provided that the initial coordinates of random walks are chosen randomly to be at or within a fixed distance of a trap, we found that the survival probability $f(t)$ first decays according to stretched exponential kinetics $f(t)=\exp(-\gamma\, t^\alpha)$, followed by a transition to power law decay $f(t)\sim t^{-\alpha}$ at longer time scales, with the same exponent $0<\alpha<1$ for both regimes. The crossover time $t_0$ between the two stages is related to the traps' absorption rate $\gamma$ by $t_0\sim \gamma^{-1/\alpha}$ and shrinks to zero in the limit of perfect traps $\gamma\to\infty$. Thus, for strongly absorbing traps the stretched exponential regime is practically absent and $f(t)$ decays according to a power law for all time scales (except very short ones). On the other hand, for weakly absorbing traps (small $\gamma$), both regimes are distinctly present. The stretched exponential regime is relatively short, lasting only while $f(t)$ has not dropped too far (roughly, less than ten per cent) from its initial value. However, the absolute duration of the regime $t_0\sim \gamma^{-1/\alpha}$ may be significant for a sufficiently small absorption rate $\gamma$. A particularly interesting instance of this case is when absorption is controlled by a thermally activated reaction, $\gamma\sim \exp(-\Delta E/k_BT)$. For a high activation energy $\Delta E\gg k_BT$, the duration of the stretched exponential regime may hold over several orders of magnitude in time.

We will show that the two-stage long-tailed kinetics of imperfect correlated traps are consistent with known results for regular spatial trap distributions~\cite{Kenkre}, and are  different from the most-studied model of perfect uncorrelated traps distributed randomly with relative concentration $0<c<1$. For that model, a good approximation for short time scales is given by the Rosenstock mean-field expression~\cite{Ros,Blumen,Blumen2}
\begin{eqnarray}
f(t)=\left\langle (1-c)^{n(t)}\right\rangle\approx(1-c)^{\langle n(t)\rangle },
\label{ros}
\end{eqnarray}
where $n(t)$ is the number of distinct sites visited by a walker. Let $r(t)$ denote the root-mean-square displacement of the walker $r(t)\sim t^{1/d_w}$, where $d_w$ is the dimension of the random walk. One recovers stretched exponential relaxation of $f(t)$ from (\ref{ros}) whenever exploration is \emph{compact}, i.e. all sites within radius $r(t)$ of the origin are visited with equal probability. In that case
\begin{eqnarray}
\langle n(t)\rangle\sim r(t)^d\sim t^{d/d_w}=t^{d_s/2},
\label{visits}
\end{eqnarray}
where $d$  and $d_s=2d/d_w$ are  the space fractal and spectral dimensions, respectively. Then it follows from (\ref{ros}) and (\ref{visits}) that the survival probability has the stretched exponential form $f(t)=\exp(-\lambda\,t^\alpha)$ with $\lambda\sim|\ln(1-c)|$ and stretching exponent $\alpha=d/d_w=d_s/2$.

At long time scales, instead of the power law kinetics we found for fractally correlated traps, the models with \emph{uncorrelated} perfect traps predict stretched exponential decay. This time, it is due to a more subtle mechanism related to rare spatial fluctuations in the trap distribution. Due to the presence of arbitrarily large trap-free regions where the walker can survive for a long time, the survival probability decays slower than exponentially, following a stretched exponential function  with  stretching exponent $\alpha=d/(d+2)$ for Euclidean host lattices~\cite{Havlin_book,Balagurov,Donsker,Grass,KRN_book} and $\alpha=d_s/(d_s+2)$ for fractals~\cite{Blumen,Blumen2,Havlin_book}. The same asymptotic behavior was also found to hold for randomly distributed {\it imperfect} traps~\cite{Nie,Tauber}. 
While some authors argue that stretched exponential long-time decay due to this mechanism occurs only when $f(t)$ drops to extremely low values and thus is not practically observable, more recent simulations show that the mechanism may manifest itself in an experimentally observable range of values of $f(t)$~\cite{Henk}. 

Thus, one will observe that the survival kinetics of uncorrelated and fractally correlated traps are qualitatively different, for both short and long time scales. Nevertheless, we shall show that the asymptotic behavior of the survival probability $f(t)$ for the model with fractally correlated traps can be accounted for by simple heuristic arguments based on the concept of compact exploration, i.e. in a manner not so different from that outlined above for the case of uncorrelated traps.

The layout of the paper is as follows. In Section 2 we formulate the model above in detail and use heuristic arguments to predict that initial relaxation for weakly absorbing traps takes a stretched exponential form. In Sections 3 through 6 we verify this prediction for a number of specific systems and discuss relevant simulation techniques, still focusing on the initial relaxation stage in systems with weak traps. The crossover to power law relaxation for longer time scales and strongly absorbing traps are discussed in Section 7. Some concluding remarks appear in Section 8.

\section{Stretched exponential kinetics}

Consider a classical particle taking discrete steps at a constant hopping rate $w$ between nearest-neighbor sites of a host lattice $\mathcal L$ of dimension $d$, with static imperfect traps interspersed according to a sublattice ${\mathcal L}_a$ of dimension $d_a \le d$. We characterize the trapping imperfection by a finite \emph{absorption rate} $w_a$, at which the survival probability decays on each trap. This induces the following master equation for $f_i(t)$, the probability that the particle is at site $i$ at time $t$:
\begin{eqnarray}
    \frac{d f_i}{dt}=w\sum_{j\in \mathcal O_i} (f_j-f_i)-
    w_a\sum_{k\in {\mathcal L}_a}\delta_{ik}\,f_i,
    \label{master}
\end{eqnarray}
where $\mathcal O_i \subset \mathcal L$ is the set of immediate neighbors of $i$.
We shall assume that the \emph{coordination number}, i.e. the number of neighbors $z=|\mathcal O_i|$, is the same for all sites $i \in \mathcal L$ of the host lattice, except those on the boundary. In this case, the master equation takes the form
\begin{eqnarray}
    \frac{d f_i}{dt}=w\sum_{j\in \mathcal O_i} f_j -z\,w\,f_i-
    w_a\sum_{k\in {\mathcal L}_a}\delta_{ik}\,f_i.
    \label{master22}
\end{eqnarray}
The problem is then to find the survival probability
\begin{eqnarray}
f(t)=\sum_{i\in \mathcal L} f_i(t).
\label{f}
\end{eqnarray}

The two characteristic parameters of the problem are the ratio $\gamma$  of the absorption and hopping rates, and the the probability $p_a$ that a walker arriving at a trap will be absorbed: 
\begin{eqnarray}
    \gamma=\frac{w_a}{w},  \qquad p_a=\frac{w_a}{z \, w+w_a}=\frac{\gamma}{z+\gamma}.
    \label{gamma}
\end{eqnarray}
The above expression for $p_a$ can be obtained as the probability that absorption occurs before the particle can hop to a neighboring site
$p_a=\int_0^\infty e^{-(z\,w+w_a)\,t}\,w_a\,dt$, 
where the exponential term  gives the (Poissonian) probability that the particle neither leaves the trap nor is absorbed over the interval $(0,t)$, and $w_a\,dt$ is the probability that absorption occurs in the interval $(t,t+dt)$.  
The limits of weak and strong absorption correspond to $\gamma,\,p_a\ll 1$ and $\gamma\gg 1$, $p_a\to 1$, respectively.

In the simulation we approximate the process
described by (\ref{master22})
by averaging over discrete random walks 
with time increment $\Delta t=1/(zw)$
and jumping probability $p=1/z$. Upon arrival at a trap site,
the particle is, in its next step, either annihilated with
probability $p_a$, or jumps to one of its $z$ neighboring sites, each with probability $(1-p_a)\,p$.

In some special cases the master equation (\ref{master22}) is amenable to analytic treatment, particularly in the continuous limit~\cite{Kenkre}. However for the general case we can glean some insight by using qualitative mean-field arguments, briefly mentioned in the previous section. 
First, summing over $i$ one obtains  from (\ref{master}) 
\begin{eqnarray}
\frac{d}{dt} f(t)=-w_a\, f_a(t), \quad f_a(t)=\sum_{k\in \mathcal L_a} f_k(t),
\label{exact}
\end{eqnarray}
where $f_a(t)$ is the probability that the particle survived up to moment $t$,
and occupies a trap at that moment. 
It may be instructive (particularly for the purpose of simulation) to define the probabilities $f(t)$ and $f_a(t)$
explicitly for an  ensemble of
$N$ particles,
\begin{eqnarray}
f(t)=\lim_{N\to\infty}\frac{N_{s}(t)}{N},\qquad
f_a(t)=\lim_{N\to\infty}\frac{N_{sa}(t)}{N},
\end{eqnarray}
where 
$N_s(t)$  is the number of particles that survived up to moment $t$, and $N_{sa}(t)$ the number of 
survivors which at that moment occupy a trap.
The above expression for $f_a(t)$ can be also presented as
\begin{eqnarray}
f_a(t)&=&\lim_{N\to\infty}\frac{N_{sa}(t)}{N_s(t)}\,\,
\frac{N_{s}(t)}{N}\nonumber\\
&=&\lim_{N\to\infty}\frac{N_{sa}(t)}{N_s(t)}\,\,
\lim_{N\to\infty}\frac{N_{s}(t)}{N}.
\end{eqnarray}
Therefore
$f_a(t)$ can be written in the form
\begin{eqnarray}
f_a(t)=P(t)\,f(t),
\label{factorization}
\end{eqnarray}
where the function  
\begin{eqnarray}
P(t)=\lim_{N\to\infty}\frac{N_{sa}(t)}{N_s(t)}
\label{P_exact}
\end{eqnarray}
has the meaning of the conditional probability that a particle that survived
up to moment $t$ occupies a trap.
From (\ref{exact}) and (\ref{factorization})
one obtains for the survival probability $f(t)$ the  
equation
    \begin{eqnarray}
        \frac{d }{dt}\,f(t)=-w_a\,P(t)\,f(t),
        \label{f}
    \end{eqnarray}
which is still exact  but of little help 
unless  one knows  the function $P(t)$ in an explicit form,
or its relation to $f(t)$.

Definition (\ref{P_exact})  suggests a straightforward way  
to evaluate $P(t)$ in a simulation
for any absorption rate; we shall briefly discuss the results of such evaluation at the end of Section 7. But in fact, for a system with weakly absorbing traps $\gamma,\,p_a\ll  1$, we can get a simple analytical approximation of $P(t)$ by speculating that at sufficiently small time scales one can neglect the effects of absorption on the occupation of traps:
\begin{eqnarray}
P(t)\approx P_0(t)=\lim_{w_a\to 0}\lim_{N\to\infty}\frac{N_{sa}(t)}{N_s(t)}=\lim_{N\to\infty}\frac{N_{a}(t)}{N}
\label{P_approx}
\end{eqnarray} 
where $N_a(t)=\lim_{w_a\to 0} N_{sa}(t)$ is the number of particles occupying traps when the absorption rate is zero. 
In other words, the approximation $P_0(t)$
is the probability of occupying a site on $\mathcal L_a$ in a system with absorption turned off.

An explicit form of the function  $P_0(t)$ is easy to 
construct for lattices of
spectral dimension $d_s< 2$  using a compact exploration argument. We first suggest that 
\begin{eqnarray}
P_0(t)=\langle n_a(t)\rangle/\langle n(t)\rangle,
\label{P1}
\end{eqnarray}
where $\langle n(t)\rangle$ and $\langle n_a(t)\rangle$ are the average  numbers of distinct sites
visited by the particle, in $\mathcal{L}$ and $\mathcal{L}_a$ respectively, up to time $t$ in a system with 
$w_a=0$.
If $d_s=2d/d_w < 2$ (or $d<d_w$) then we ensure that diffusive exploration is
compact, and therefore $\langle n_a(t)\rangle$ and $\langle n(t)\rangle$
approximate the average number of sites in $\mathcal L_a$ and $\mathcal L$ within a radius
$r(t)
\sim (wt)^{1/d_w}$, i.e.~the root-mean-square
displacement of the particle on a trap-free lattice. This gives
    \begin{eqnarray}
        \langle n(t)\rangle &\sim& r(t)^d\sim(wt)^{d/d_w},\nonumber\\
        \langle n_a(t)\rangle &\sim& r(t)^{d_a}\sim(wt)^{d_a/d_w}.
        \label{aux111}
    \end{eqnarray}
Substitution of (\ref{aux111}) into (\ref{P1}) 
gives for  $P_0(t)$ an  asymptotic  
power law
    \begin{eqnarray}
        P_0(t)\sim\,(w\,t)^{-\beta},\quad \beta=\frac{d-d_a}{d_w}.
        \label{P0}
    \end{eqnarray}
Then the corresponding solution of
(\ref{f})
    \begin{eqnarray}
        f(t)=\exp\left\{-w_a\,\int_0^t P_0(t')dt'\right\}
        \label{f2}
    \end{eqnarray}
has stretched exponential form
    \begin{eqnarray}
        f(t)=\exp\left\{-c\,\gamma\,(w\,t)^\alpha\right\},
        \label{f3}
    \end{eqnarray}
with $\alpha=1-\beta$,
    \begin{eqnarray}
        \alpha=1-\frac{d-d_a}{d_w}.
        \label{alpha}
    \end{eqnarray}
The empirical constant $c$ in (\ref{f3})
remains undefined, but is expected to be of order of one.
The experimental evidence of such a relaxation would be a linear dependence of $\ln(-\ln(f))$ versus $\ln t$, with slope $\alpha$.

In order to facilitate  comparison with discrete time simulation, it is convenient to express $\gamma$ in terms of the absorption probability $p_a$. Rearranging (\ref{gamma}), we obtain $\gamma=z\,p_a/(1-p_a)$. By also taking into account that the time unit in our simulation is $1/zw$, we may then
write (\ref{f3}) as
 \begin{eqnarray}
        f(t)=\exp\left\{-c\,\frac{z\,p_a}{1-p_a}\,(z\,w\,t)^\alpha\right\}.
        \label{f3M}
    \end{eqnarray}
The empirical constants $c$ in Eqs. (\ref{f3}) and (\ref{f3M})
differ by a factor of $z^\alpha$.


Although compact exploration only holds for Euclidean lattices when $d=1$, we shall see in Section 6 that our approach still produces a reasonable approximation when $d=2$. On the other hand, many types of fractals satisfy the condition of $d_s < 2$. In particular, for random walks on a critical percolation  cluster, it holds for any dimension of the embedding lattice~\cite{Havlin_book}. It is not \emph{a priori} clear how to extend the above reasoning to the case where $d_s > 2$, and exploration is no longer compact.
We leave this as an open question, and shall not discuss it below.

One expects that the above mechanism of stretched exponential relaxation is limited in both short and long time
scales. On one hand, the mean-field-like expression (\ref{P1}) and scaling
relations (\ref{aux111}) presuppose that the walker has performed many steps,
$wt \gg 1$; the simulations below give an empirical lower bound of
between 10 and 100 steps. On the other hand, the above estimation of $P(t)$ 
assumes
that the visiting frequency of trapping sites is not affected by annihilation,
which implies that $f(t)$ must be close to one. From (\ref{f3}) one estimates
the upper bound to be $wt\ll\gamma^{-1/\alpha}$. Thus the validity domain of the
proposed mechanism for an infinite system is expected to be
    \begin{eqnarray}
        1 \ll wt\ll \gamma^{-1/\alpha}.
        \label{validity}
    \end{eqnarray}
For a finite system of size $L$, the upper bound is a minimum of $t_0\sim \gamma^{-1/\alpha}$ and $t_1\sim L^{d_w}$. We stress again that while the interval (\ref{validity}) corresponds only to the initial stage of relaxation, the absolute duration of this stage for weakly absorbing traps
($\gamma\ll 1$) may be significant.

For times much larger than $t_0=w^{-1}\,\gamma^{-1/\alpha}$, the approximation (\ref{P0}) for $P(t)$ ceases to be valid, and simulation shows that stretched exponential relaxation is replaced by power law decay $f(t)\sim t^{-\alpha}$ with the same $\alpha$ given by (\ref{alpha}). We shall postpone detailed discussion of this regime until Section 7.

The previous argument relies on the assumption that traps are weakly absorbing.
For strong absorption rates $\gamma\agt 1$, including the limit of perfect traps $\gamma\to\infty$, the validity interval (\ref{validity}) of the stretched exponential regime is inconsistent and, as we shall see, the approximation
(\ref{P_approx})
for $P(t)$ is not valid at any time. In this case, as we discuss in Section 7, the stretched exponential regime is absent and $f(t)$, after a short transient period, follows the same power law as for weakly absorbing traps at long time scales.

Another restriction on our heuristic argument for the stretched exponential decay of $f(t)$ is that the initial location of the walker must be on, or within a fixed distance of, a trap. Otherwise (e.g. if an initial site is chosen randomly on the host lattice) the second asymptotic relation in (\ref{aux111}) may be invalid. If the initial distance between the particle and a trap is $i_0$ (in lattice spacing units), then one will expect that the above reasoning starts to work only after the time needed for a walker to reach a trap, $wt\sim i_0^{d_w}$. In this case, instead of (\ref{validity}) one expects a validity interval with a higher lower bound, namely
    \begin{eqnarray}
        i_0^{d_w} \ll wt\ll \gamma^{-1/\alpha}.
        \label{validity_M}
    \end{eqnarray}\
Hence for the validity interval to be significant we require that $i_0\ll \gamma^{-\frac{1}{\alpha\,d_w}}$, which again can only be a meaningful condition when absorption is weak, $\gamma \ll 1$. 

Note that mathematically our model is similar to the class of defect-diffusion
models of dipole relaxation, which assume that dipole reorientation (relaxation)
is triggered by mobile defects~\cite{Montroll,Klafter,Kaka,Bendler}. In this
case the rate equation for the fraction of surviving dipoles $f(t)$ has the form
(\ref{f}), where $P(t)$ is now the diffusive current of defects. If the latter
is characterized by power law decay like (\ref{P0}), then this model is formally
equivalent to ours. The two models do, however, employ very different mechanisms
to argue power law decay of $P(t)$. In our model it is due to fractal spatial
correlations of traps, whereas in defect-diffusion models it is due to
dispersive transport of defects (in which case $\alpha$ is typically
temperature-dependent). The difference is also reflected in the fact that the
validity range of defect-diffusion models is not restricted by the initial time interval and, when relevant, is capable of describing a much larger
section 
of the relaxation function than the model we discuss here.

As a final note for this section, we assumed above
that the exponent $\beta$ in (\ref{P0}) is less than one. For the special case
where $\beta=1$, the above
reasoning leads, instead of to stretched exponential decay, to a power law
$f(t)=(wt)^{-c\,\gamma}$ with an empirical constant $c$.


\section{1D lattice with a single trap}
As a simple test of the qualitative argument  outlined in the previous section,
consider the limiting setting when the host lattice  is one-dimensional,
diffusion is regular, and the trapping  sublattice consists of a single trap
site:
    \begin{eqnarray}
        d=1,\quad d_w=2,\quad d_a=0, \quad d_s=1.
    \end{eqnarray}
For a weakly absorbing trap  we expect the initial relaxation of the survival probability to follow a
stretched exponential law with
$\alpha=1-(d-d_a)/d_w=1/2$.

If the trap site is at position $i=0$, then the master equation (\ref{master22})
has the form
    \begin{eqnarray}
        \frac{d f_i}{dt}=w\,f_{i-1}+w\,f_{i+1}-2w\,f_i-w_a\,\delta_{i0}\,f_i
        \label{master2}
    \end{eqnarray}
with the initial condition $f_i(0)=\delta_{ii_0}$ that the particle starts at the initial site $i_0$, which may coincide with the trap if $\gamma = w_a/w$ is finite. Eq. (\ref{exact}) 
takes the form $df/dt=-w_a\,f_0$, and 
the survival probability is determined by the probability to be on the trap site $f_0(t)$:
    \begin{eqnarray}
        f(t)=1-w_a\,\int_0^t f_0(t)\,dt.
    \end{eqnarray}
This problem is exactly solvable in the continuous limit (see~\cite{Kenkre} and references therein). 
We outline the solution in the Appendix and show that over interval (\ref{validity_M}), which in this case becomes
    \begin{eqnarray}
        i_0^2 \ll wt\ll \gamma^{-2},
        \label{validity2}
    \end{eqnarray}
the survival probability has the approximate form 
    \begin{eqnarray}
        f(t)\approx 1-c\,\gamma\, \sqrt{w\,t}
        \label{analytic}
    \end{eqnarray}
with $c=1/\sqrt{\pi}\approx 0.56$.
This is consistent with prediction (\ref{alpha}): over interval
(\ref{validity2}) for small $\gamma$, expression (\ref{analytic}) is a good approximation of the the stretched exponential function with $\alpha=1/2$,
    \begin{eqnarray}
        f(t)=e^{-c\,\gamma\,\sqrt{wt}}=
        \exp\left\{-\frac{c\,\sqrt{2}\,p_a}{1-p_a}\,\sqrt{2\,w\,t}\right\}.
        \label{ser2}
    \end{eqnarray}

\begin{figure}
  \includegraphics[height=6.2cm]{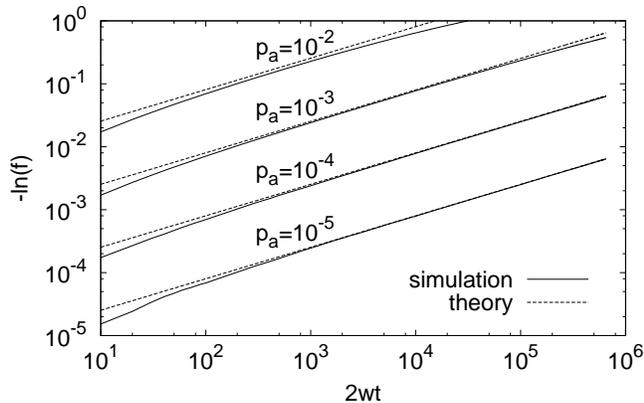}
  \caption{A log-log plot of the function $-\ln(f(t))$
  for random walks on a line
  with a single trap for different values of the absorption probability $p_a$. For
  all curves, the initial position of the walker coincides with the position of
  the trap. Solid lines show the simulation results (averaged over about $10^7$
  trajectories) and dashed lines show corresponding stretched exponential
  curves according to Eq. (\ref{ser2}) with $c=0.56$.}
  \label{fig_1}
\end{figure}

\begin{figure}
  \includegraphics[height=6.2cm]{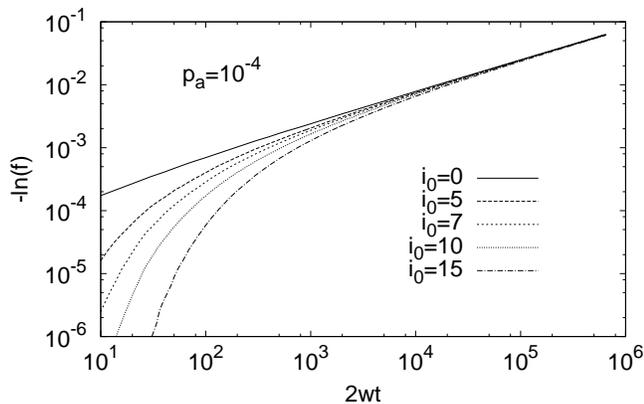}
  \caption{A log-log plot of simulation data for the function $-\ln(f(t))$
  for random walks on a line with absorption probability $p_a=10^{-4}$
  and different initial positions $i_0$. A single trap is located at the origin,
  $i=0$.}
  \label{fig_2}
\end{figure}

We found the result to be in good agreement with numerical simulation. A comparison of (\ref{ser2}) with a numerical experiment for different (small) values of the absorption probability $p_a$ 
and initial position $i_0=0$ is presented in
Fig. 1. 
According to (\ref{ser2}), 
a log-log plot of the function 
$-\ln(f(t))$, i.e. 
the plot of $\ln(-\ln(f))$ versus $\ln(t)$,
must appear to be a straight line with slope given by the exponent $\alpha=1/2$.
Our simulation confirms this prediction, and shows an increase in the duration of its validity domain 
for smaller $p_a$ and  $\gamma$ in a way that is consistent with
(\ref{validity2}). All curves show deviation from stretched exponential
relaxation for small $t$ when $2wt$ is of order ten.
Deviation for large $t$ is easily noticeable for $p_a=10^{-2}$ and
$p_a=10^{-3}$, but is beyond the experiment's time range for the curves with
$p_a=10^{-4}$ and $p_a=10^{-5}$. Fig. 2 shows 
similar plots for the same
value $p_a=10^{-4}$ and different initial positions $i_0\ge 0$ of the
particle. In those cases, the stretched exponential regime emerges after 
a transient time which increases quadratically in $i_0$; this is in agreement with (\ref{validity2}).

Similar results hold
for the mathematically equivalent problem of a two-dimensional host lattice
with traps on a one-dimensional line:
    \begin{eqnarray}
        d=2,\quad d_w=2,\quad d_a=1, \quad d_s=2.
    \end{eqnarray}
In this case, equation (\ref{alpha}) predicts stretched exponential
relaxation with $\alpha=1/2$, and numerical simulation confirms this over
the time interval (\ref{validity2}).
The simulations also show agreement
with theoretical predictions for other types of
simple (Euclidean) trap configurations
embedded in one- and two-dimensional host lattices. In the sections to 
follow, we consider host and trap lattices with nontrivial fractal dimensions.

\begin{figure*}[t]
    \includegraphics[height=5.5cm]{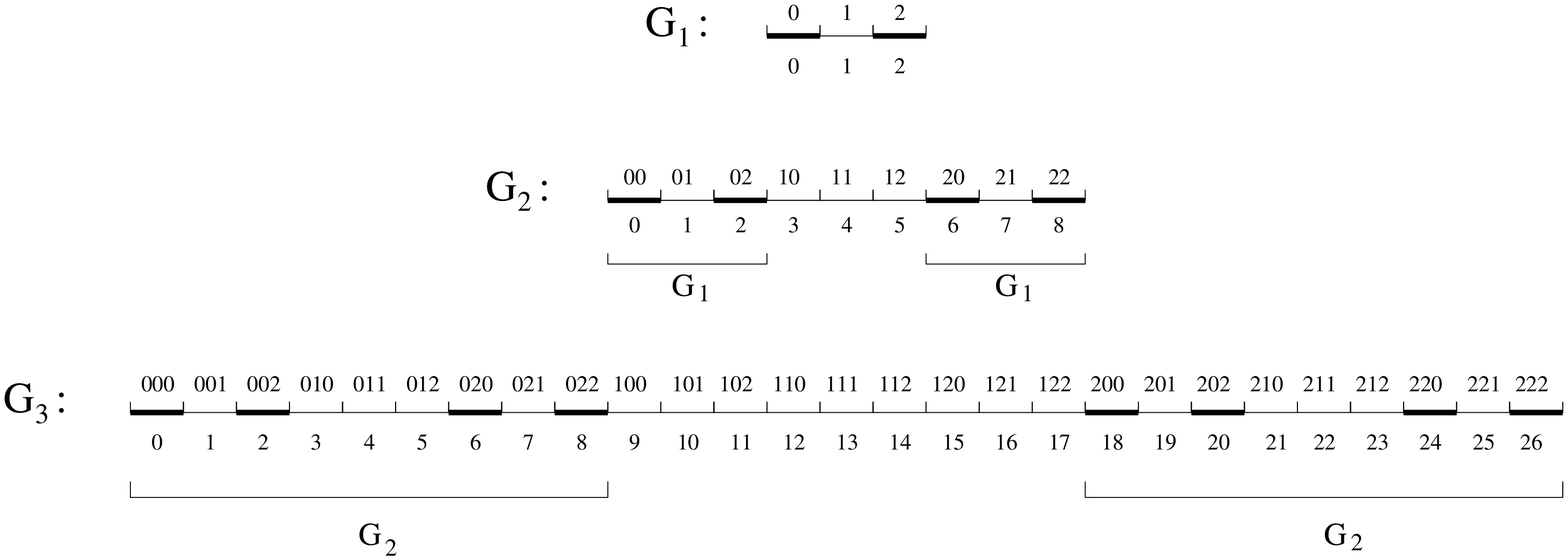}
    \caption{An iterative construction of a 1D lattice, with trap sites
    (highlighted by bold segments) forming a Cantor set. The first three
    generations $G_1, G_2$, and $G_3$ are shown.
    The cells of the host lattice are enumerated from left to right with
    nonnegative integers. Below each cell is its corresponding label in decimal
    (base-10) notation, and above is that same integer's ternary (base-3)
    representation. A site is a trap if and only if its ternary label does
    not contain the digit `1'.}
\end{figure*}


\section{1D lattice with traps on the Cantor set}

We shall now extend the previous example to have traps
distributed along a constructive analogue of the Cantor set. We shall
refer to the host lattice, together with the traps, as the Cantor lattice.
In order to simplify the enumeration of trapping sites, in this and the following sections we shall associate hopping sites with lattice cells, rather than the cells' end points.

Unlike the Cantor set, which is defined ``from
the outside, in'' by starting from the (uncountable) interval $[0,1]$ and
recursively removing the middle third of every resulting subinterval, we employ
an ``inside-out'' (and countable) iterative construction; see Fig. 3. The first
generation $G_1$ consists of three consecutive cells, the first and last of
which are traps. Then we define the $n$-th generation recursively to be two
copies of $G_{n-1}$ flanking a trap-free $G_{n-1}$-sized block. Hence
each $G_n$ has size $3^n$ and contains $2^n$ traps.
One refers to each generation $G_n$ as a \emph{finite} Cantor lattice, 
and the limit $n \to \infty$ as \emph{the}
Cantor lattice,
characterized by the following dimensions:
    \begin{eqnarray}
        d=1,\quad d_w=2,\quad d_a=\ln 2/\ln 3, \quad d_s=1.
    \end{eqnarray}
According to the heuristic argument in Section 2, the survival probability for weakly absorbing traps is
expected to be characterized by stretched exponential decay (\ref{f3}) with the exponent
    \begin{eqnarray}
        \alpha=1-\frac{d-d_a}{d_w}=\frac{1}{2}\left(
        1+\frac{\ln 2}{\ln 3}
        \right)\approx 0.8155.
        \label{alpha3}
    \end{eqnarray}

We verified this prediction with numerical simulations of $10^7$ random walks, 
each starting from a random trap cell, on
the finite Cantor lattice $G_{20}$, which consists of over $10^9$ cells. 
In fact one could go so far as to dynamically generate the Cantor lattice,
but averaging over initial conditions for a sufficiently large but finite
lattice like $G_{20}$ is simpler, and the 
size effects are negligible.
(We found empirically that for random  walks of about $10^6$ steps, finite-size effects become noticeable only for lattices smaller than $G_{10}$.)
In lieu of explicitly storing the
locations of traps, which would be infeasible, we will exploit a
property of the natural left-right enumeration (see Fig. 3) of any given
generation. Let $i$ be the label of a cell in $G_k$, and
$t(i) = (d_1 \dots d_k)$ its ternary representation, 
i.e. the unique
sequence of $d_1,\dots,d_k \in \{0,1,2\}$ such that
$i = d_k 3^0 + d_{k-1} 3^1 + \dots + d_1 3^{k-1}$. The convenience of this
representation is that traps can be identified immediately:
A site with a ternary address $t=(d_1d_2\cdots d_k)$ is a trap 
if and only if
$d_j \neq 1$ for all $j \le k$; see Fig. 3.  
In other words, the subset $\mathcal L_a$ of trap sites
on $G_k$ is exactly
    \begin{eqnarray}
    \mathcal L_a=\{i: t(i)=(d_1 \dots d_k) \mbox{ and }  d_j\ne 1 \mbox{ for all } j\}.
        \label{La}
    \end{eqnarray}

The relevance of a ternary enumeration is made apparent and natural when one
considers the recursive composition of the Cantor lattice: Every generation
consists of left, right, and middle parts of equal length, and only the latter
is guaranteed to be trap-free. Then one may read the ternary label $(d_1 \dots
d_k)$ as a sequence of choices: $d_1 = 0,1,2$ correspond to the left,
middle, and right $G_{k-1}$-sized blocks within $G_k$, and so on for each 
successive digit. For
example, in the $G_3$ lattice, the cell labeled $(012)$ is in the left $G_2$
block, since $d_1 = 0$. That  specific $G_2$ block is made of three $G_1$
blocks, and $d_2 = 1$ indicates that we are in the middle one. Finally, within
that specific $G_1$ block the cell is on the right, so $d_3=2$. On the other
hand, the cell has the decimal label $5$, which is precisely the decimal
representation of the ternary number $(012)$.

\begin{figure}
  \includegraphics[height=6.2cm]{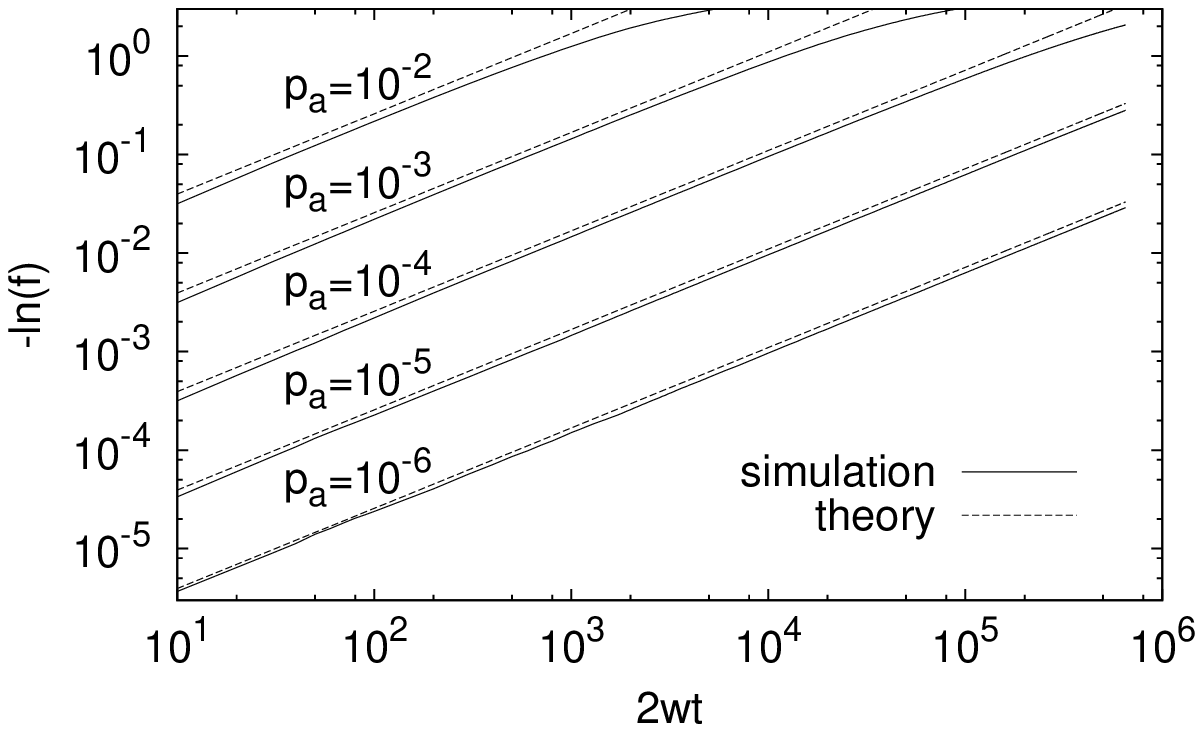}
  \caption{A log-log plot of the function $-\ln(f(t))$ for $10^7$ random walks on a
  $G_{20}$ Cantor lattice, for several values of the absorption probability $p_a$.  Solid lines
  show the simulation results, dashed lines show the corresponding stretched
  exponential curves according to Eq.~(\ref{f3M}) with $\alpha=0.8155$, as given by
  (\ref{alpha3}), the coordination number $z=2$,  and the empirical parameter $c=0.3$. Each trajectory begins
  from a randomly selected trap.}
  \label{fig_4}       
\end{figure}

\begin{figure*}[t]
  \includegraphics[height=5.5cm]{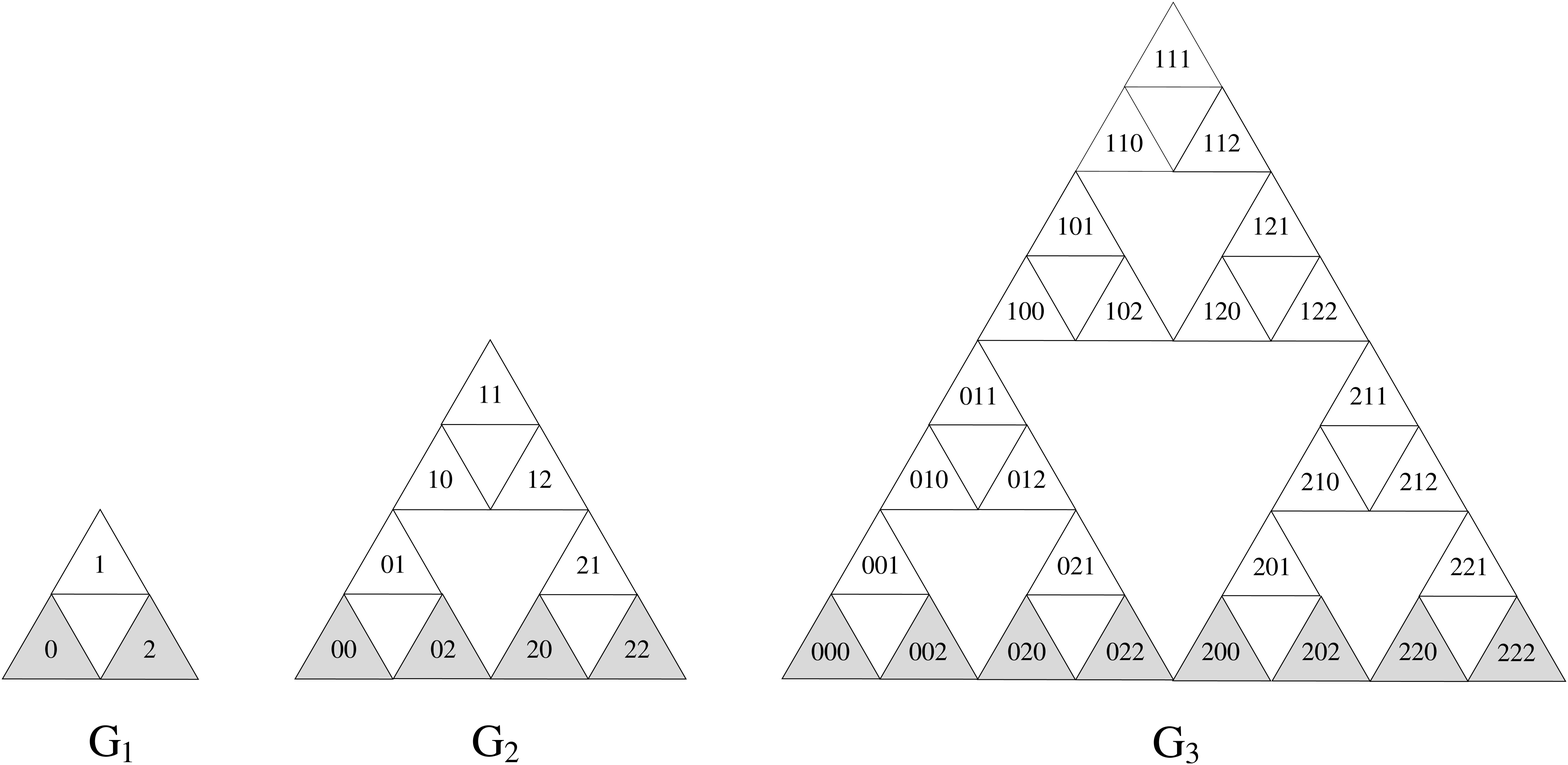}
  \caption{Composition  and site enumeration schemes  for the Sierpinski
  lattice. The first three generations $G_1$, $G_2$ and $G_3$  are shown. Traps,
  depicted as shaded elementary triangles, are characterized by ternary labels that do not contain the digit `1'.}
 \label{fig_5}
\end{figure*}



Simulation results for $G_{20}$ are shown in Fig.~4. After a transient of
about ten steps, the initial relaxation of the survival probability closely
follows stretched exponential kinetics (\ref{f3})
with the exponent $\alpha$
given by (\ref{alpha3}). One observes that the upper time bound for this behavior
increases as the absorption probability 
$p_a$ 
decreases,
in a way consistent with prediction (\ref{validity}).
As $t$ exceeds the interval of validity, the transition to slower
(power law) relaxation occurs, which in Fig. 4 is visible for
$p_a\ge 10^{-4}$. For $p_a=10^{-5}$ and
$p_a=10^{-6}$ the transition  is beyond 
the simulation's time scope.
We postpone discussion of the long-time power-law relaxation regime and systems with strong absorption until Section 7.

If the initial site were not a trap, but instead chosen to be at a given
distance $x_0$ away from a (randomly chosen) trap, then the
relaxation curves would have a form similar to that
in Fig. 2, 
i.e. approaching stretched exponential form
at long times scales, with
a transition time increasing with $x_0$.

\section{Sierpinski  lattice  with 1D set of traps}
For our next example, we will consider random walks on a constructive
variant of the (fractal) Sierpinski gasket, with traps located on
a one-dimensional subset, namely the bottom ``edge''; see Fig.~5.
This case differs from the two previous examples, in that diffusion
on the host lattice is anomalous, with $d_w>2$.
Here the set of relevant dimensions is~\cite{Havlin_book}
    \begin{eqnarray}
        d=\frac{\ln 3}{\ln 2},\quad d_w=\frac{\ln 5}{\ln 2},
        \quad d_a=1, \quad d_s=2\, \frac{\ln 3}{\ln 5}.
    \end{eqnarray}
We  expect for the initial relaxation of the survival probability
to have the stretched exponential
form (\ref{f3})
with the exponent
\begin{eqnarray}
\alpha=1-\frac{d-d_a}{d_w}=\frac{\ln (10/3)}{\ln 5}\approx0.748,
\label{alpha5}
\end{eqnarray}
as long as the starting point of each random walk is on or near a trap and absorption is weak $p_a\ll 1$.
Numerical simulation
confirms this prediction with time bounds
similar to those for the two previous cases; see Fig.~6.
Below we discuss some technical details of the simulation,
which for Sierpinski lattices has some
peculiarities of its own.

\begin{figure}
  \includegraphics[height=6.2cm]{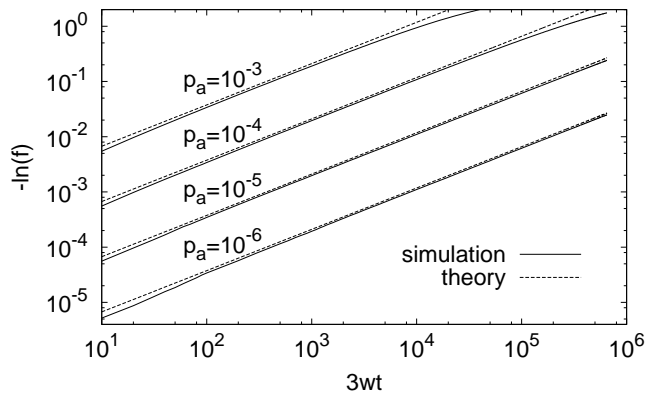}
  \caption{A log-log plot of the function $-\ln(f(t))$ for random walks on
the Sierpinski lattice $G_{20}$ with a one-dimensional
sublattice of traps (see Fig. 5)
for different values
of the absorption probability $p_a$.
Solid lines show the simulation results
(averaged over about $10^7$ trajectories), dashed lines show
the corresponding stretched exponential curves according
to (\ref{f3M}) with $\alpha=0.748$, as given by (\ref{alpha5}),
the coordination number of traps $z=3$,
and the empirical constant $c=0.4$. Initial sites are chosen randomly
from the trap lattice $\mathcal L_a$.}
  \label{fig_6}       
\end{figure}

For the same reasons as the Cantor lattice, it is convenient to
generate our Sierpinski lattice using a recursive ``inside-out'' construction. This induces a natural ternary enumeration of the lattice cells, depicted in Fig. 5.
As in the previous section, the hopping sites of the walk are the cells of the lattice, which this time are the ``elementary'' triangles of every generation $G_k$.
The lattice of the first generation $G_1$ consists of three triangles
whose positions are labeled $0$ (left), $1$ (top), and $2$ (right).
The lattice of the second
generation $G_2$ consists of three $G_1$ blocks, whose three positions
``left'', ``top'', and ``right'' are again
denoted $0$, $1$, and $2$ respectively. The three $G_2$ blocks compose
in a similar manner to make 
the third generation lattice $G_3$, and
the process may be repeated to any desirable order.
Hence we may once again use the ternary addressing scheme wholesale,
and the traps are exactly those sites without any digits equal to 1.
Indeed the only difference, from the perspective of simulation, between
this and the preceding section is that the set of neighbors for each
cell has changed.
As in the preceding section, 
the simulation was carried out on the
lattice $G_{20}$, consisting  of $3^{20}\sim 10^9$ cells,
which we found to be large enough for finite-size effects
to be negligible.


 \begin{figure*}[t]
\includegraphics[height=6.6cm]{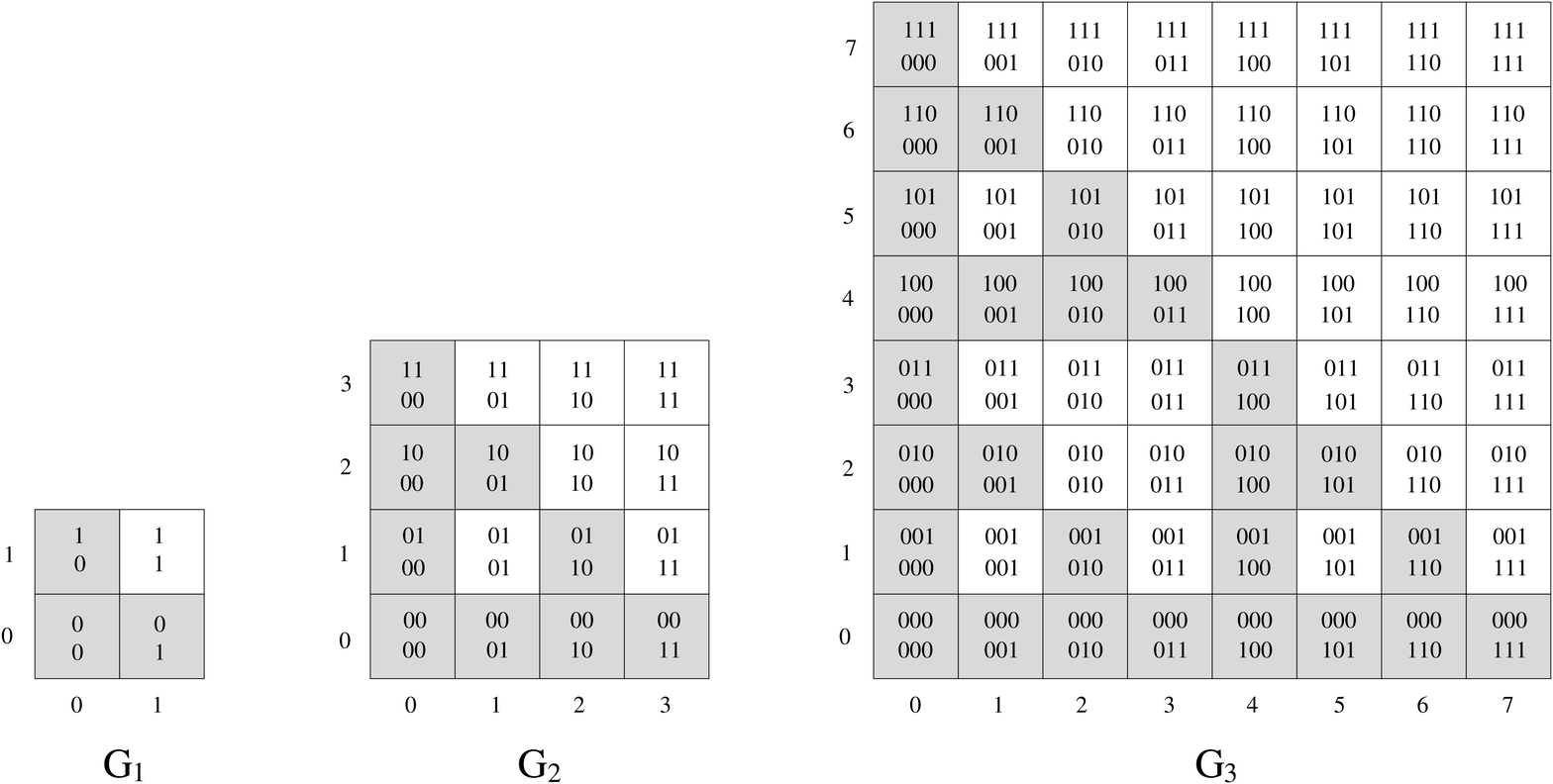}
\caption{Recursive composition  and cell enumeration of a 2D lattice with traps
on the Sierpinski gasket (depicted by shaded cells). The first three generations $G_1, G_2, G_3$ 
are shown. 
Each cell is decorated with its binary address, which is the pair of its Cartesian coordinates $x$ (bottom) and $y$ (top), represented in base 2.}
 \label{fig_7}
\end{figure*}

As we foreshadowed,
in order to simulate random walks on a Sierpinski lattice
of generation $k$, one needs an algorithm for determining
the neighbors of a given cell. 
Suppose the cell has label
    \begin{eqnarray}
        t_0=(d_1\cdots d_{k-1}\,d_k).
        \label{label}
    \end{eqnarray}
Two of its neighbors (both for an apex cell) must belong to the same
$G_1$-block as $t_0$, meaning the neighbors' labels $t_1$ and $t_2$ differ
from $t_0$ only by the final digit:
\begin{eqnarray}
\!\!\!\!\!\!\!\!\!\!\!\!\!\!\!\!\!
&&t_1=(d_1\cdots d_{k-1}\, d'_k), \quad d_k'=(d_k+1)\bmod 3,\nonumber\\
\!\!\!\!\!\!\!\!\!\!\!\!\!\!\!\!\!
&&t_2=(d_1\cdots d_{k-1}\, d''_k), \quad d_k''=(d_k+ 2)\bmod 3.
\label{label0}
\end{eqnarray}
For example, the cell at $(012)$ in the $G_3$ lattice has two neighbors
from the same $G_1$-block, with labels $(010)$ and
$(011)$; see Fig. 5.

However, the algorithm for finding the label $t_3$ of the third neighbor is more
involved~\cite{Rammal,Weber}. First, for a given cell with the label
(\ref{label}), one checks whether $d_k\ne d_{k-1}$. If the condition is
satisfied, i.e the label has the form
    \begin{eqnarray}
        t_0=(d_1\cdots d_{k-2}\,\, y\,\, z), \quad y\ne z,
        \label{label2}
    \end{eqnarray}
then the cell and its third neighbor belong to different $G_1$ blocks but to the
same $G_2$ block. In this case the label of the third neighbor $t_3$  has the
same first $k-2$ digits as $t_0$, while the last two digits replace
each other:
    \begin{eqnarray}
        t_3=(d_1\cdots d_{k-2}\,\,z\,\, y).
    \end{eqnarray}
For example, for the cell with $t_0=(012)$, the third neighbor is labeled
$t_3=(021)$; see Fig. 5.

Now, if the condition above were not satisfied, i.e.~$d_k=d_{k-1}$, then one
must check whether $d_{k-1}\ne d_{k-2}$, i.e.
\begin{eqnarray}
t_0=(d_1\cdots d_{k-3}\,\, y\,\, z\,\,z), \quad y\ne z.
\label{label3}
\end{eqnarray}
In this case the cell and its third neighbor belong to different $G_1$ and $G_2$
blocks, but to the same $G_3$ block. Then the label of the third neighbor $t_3$
has the same first $k-3$ digits as the label $t_0$ (\ref{label3}),  while the
last three digits $yzz$ are replaced by $zyy$,
    \begin{eqnarray}
        t_3=(d_1\cdots d_{k-3}\,\,z\,\, y\,\,y).
    \end{eqnarray}
For example, in $G_3$ the cell with $t_0=(011)$ has its third neighbor labeled
$t_3=(100)$; see Fig.~5.

One will by now see the pattern of this method. Proceed as
above until meeting the condition that
$d_{k-i}\ne d_{k-i-1}$, in which case
\begin{eqnarray}
t_0=(d_1\cdots d_{k-i-2}\,\, y\,\, \underbrace{z\,\,z \cdots z}_{\text{$i$ digits}}),\quad y\ne z.
\label{label4}
\end{eqnarray}
Then the label of the third neighbor has the form
\begin{eqnarray}
t_3=(d_1\cdots d_{k-i-2}\,\, z\,\, \underbrace{y\,\,y \cdots y}_{\text{$i$ digits}}).
\label{label5}
\end{eqnarray}
Of course, if all the digits $d_1 = \dots = d_k$ are equal, then $t_0$ labels an apex cell, and there is no third neighbor.

For example, in $G_5$ of the Sierpinski lattice the cell with $t_0=(10222)$ has
two neighbors with labels given by
(\ref{label0}), namely $t_1=(10220)$ and $t_2=(10221)$, and the third
neighbor with the label $t_3=(12000)$  given by (\ref{label5}) (with $i=3$).

The construction, enumeration, and nearest-neighbor algorithms
outlined in this section
can be readily extended to Sierpinski lattices
of higher dimensions.

\section{2D lattice with traps on Sierpinski gasket}
For our final example, we consider random walks 
on a two-dimensional Euclidean lattice with
traps forming a Sierpinski gasket; see Fig. 7. In this case,
    \begin{eqnarray}
        d=2,\quad d_w=2,
        \quad d_a=\frac{\ln 3}{\ln 2}, \quad d_s=2,
    \end{eqnarray}
and according to Sec.~2 the stretching exponent should be
    \begin{eqnarray}
        \alpha=1-\frac{d-d_a}{d_w}=\frac{1}{2}\,\frac{\ln 3}{\ln 2}\approx 0.79.
        \label{alpha6}
    \end{eqnarray}
One might worry more about the validity of this prediction than for our other
examples because, strictly speaking, a random walk in two dimensions is not
compact: The average number of distinct visited sites 
$\langle n(t)\rangle$
increases in 2D as $t/\ln t$  \cite{KRN_book,Havlin_book}, rather than linearly,
as anticipated by the compact-exploration ansatz (\ref{aux111}). Yet one may
expect that the slowly varying logarithmic factor can be approximated without much error
by a constant, so that the prediction (\ref{alpha6}) may still be justified.
Numerical simulation supports this optimism: The slope of $f(t)$ in the
double-logarithmic scale is found to be in good agreement with (\ref{f3M}) and
(\ref{alpha6}); see Fig.~8.

Similarly to the previous two sections, in our simulation we construct the 2D
lattice with the embedded Sierpinski gasket by composing higher generations from
lower ones recursively, as shown in Fig.~7. Simulation  results presented in
Fig.~8 are for the lattice $G_{20}$, with periodic boundary
conditions. Simulations for lattices larger than $G_{10}$ and with other types
of boundary conditions show very similar results, indicating that 
finite-size effects are negligible.

\begin{figure}
  \includegraphics[height=6.1cm]{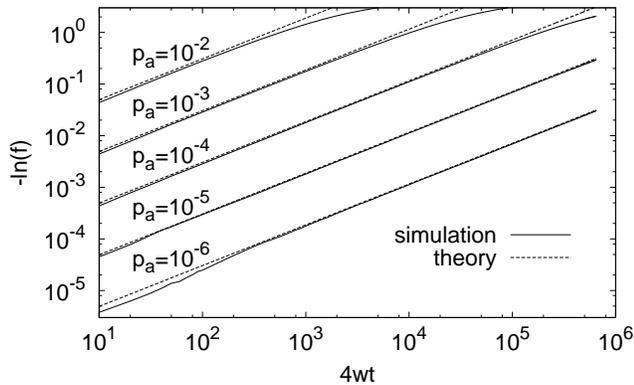}
  \caption{A log-log plot of the function $-\ln(f(t))$ for random walks on a 2D
  lattice with traps on the Sierpinski gasket  $G_{20}$ (see Fig. 7) for different values
  of the absorption probability $p_a$. Solid lines show the simulation results (averaged over
  about $10^7$ trajectories), dashed lines show the corresponding stretched
  exponential curves according to (\ref{f3M}) with $\alpha=0.79$, as  given by
  (\ref{alpha6}), the coordination number $z=4$, and the empirical constant $c=0.2$. Initial sites are chosen
  randomly from among the trap sites.}
  \label{fig_8}       
\end{figure}

Cells are enumerated by pairs of Cartesian coordinates $(x,y)$ expressed in
binary. For a lattice $G_{k}$ of generation $k$, binary coordinates have $k$
digits
    \begin{eqnarray}
        x=(a_1 a_2 \cdots a_k),\quad y=(b_1 b_2 \cdots b_k),
    \end{eqnarray}
where each digit $a_i, b_j$ has a value of either zero or one; see Fig.~7. The
binary enumeration is convenient because traps (i.e.~cells belonging to the
Sierpinski gasket) can be identified as those, and only those, cells whose
binary addresses satisfy the condition that the sum of digits in every position
does not exceed one,
    \begin{eqnarray}
        a_i+b_i<2, \mbox{ for }     i=1,2, \cdots k.
        \label{trap_condition}
    \end{eqnarray}
(In other words, the bitwise `AND' operation applied to $x$ and $y$ is zero if and only if the
site is a trap.) For example, in the lattice $G_3$ (see Fig. 7) for the trapping
cell with the binary address $x=(010), y=(101)$, we have $a_i+b_i=1$ for $i=1,2,3$, and therefore
the condition (\ref{trap_condition}) is satisfied. On the other hand, for the
non-trapping cell with binary address $x=(101)$, $y=(011)$ 
the condition (\ref{trap_condition}) is not satisfied because 
$a_3+b_3=2$.
This method can also be used to model random walks on 
the Sierpinski gasket, as a perhaps simpler alternative to the method 
described in the previous section.

\begin{figure}
  \includegraphics[height=6.1cm]{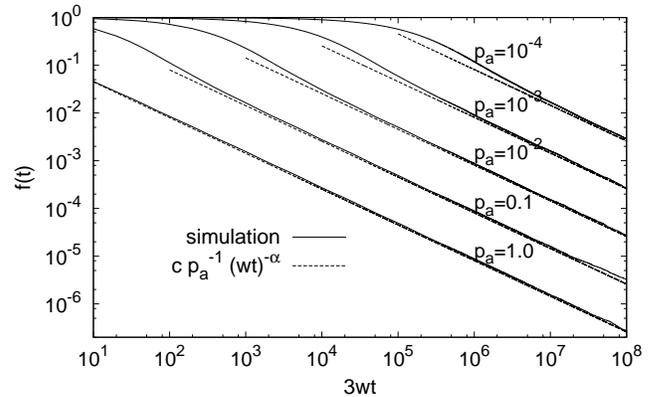}
  \caption{
  A log-log plot of the survival probability $f(t)$ for random walks on the  Sierpinski lattice $G_{30}$ with a one-dimensional sublattice of traps (see Fig. 5) for several values
  of the absorption probability $p_a$. Solid lines show simulation results (averaged over between $10^6$ and $10^8$ walks) and dashed lines show 
  the corresponding power law relaxation functions given by Eq. (\ref{power}),
  $f(t)=c\,p_a^{-1}\,(3wt)^{-\alpha}$, with $\alpha=0.748$ and empirical constant $c=0.25$. 
  }
  \label{fig_9}       
\end{figure}

\section{Power-law kinetics} 
For all the example systems discussed above, simulation shows that for weakly absorbing traps ($\gamma, p_a\ll 1$) the initial  stretched exponential kinetics are replaced at longer time scales $t\gg t_0=w^{-1}\gamma^{-1/\alpha}$ by algebraic decay with the same exponent $\alpha$ as for the stretched exponential regime, and with a prefactor proportional to the inverse absorption probability,
    \begin{eqnarray}
        f(t)\sim p_a^{-1}\,(w\,t)^{-\alpha}, \qquad 
        \alpha=1-\frac{d-d_a}{d_w}.
        \label{power}
    \end{eqnarray}
In this case, double-logarithmic plots of $f(t)$ at long times become straight lines with slope $-\alpha$. As the absorption rate increases, the crossover time $t_0$ decreases, and the stretched exponential regime becomes less visible. For strongly absorbing ($\gamma\gg 1$) and perfect ($\gamma\to \infty$, $p_a=1$) traps, $f(t)$ follows power law kinetics for all times, except very short ones. Fig. 9 illustrates this behavior for random walks on the  Sierpinski lattice described in Section 5, considering generation $G_{30}$ with initial coordinates chosen randomly from the sites adjacent to a trap. For other systems the results are similar (but curiously, the graph produced by the Cantor lattice has some small but noticeable ripples, even after averaging over a great many walks).

Fig. 9 reveals that the long-time behavior of the survival probability $f(t)$ is the same for imperfect ($p_a<1$) and perfect ($p_a=1$) traps. This is a well-known result for the case of a {\it single} trap (see Appendix and ~\cite{Havlin_book}), and we see that it persists for a network of correlated traps as well. It suggests that on a long time scale, regardless whether traps are perfect, the kinetics of absorption are controlled not by occupation statistics (as we assumed in Section 2 when evaluating $P(t)$, the probability to occupy a trap), but rather by first passage time (FPT) statistics. While for perfect traps the relevance of the FPT statistics is obvious, for imperfect traps it can be understood intuitively by speculating that the main contribution to the survival probability at long time-scales comes from particles performing long excursions in large trap-free regions.
The duration of such an excursion is essentially the time of first return to the absorbing lattice, and assumed to be much larger than the time the particle spends after returning to a trap-rich region. We show below that this picture leads naturally to the long-time asymptotic behavior of (\ref{power}).

In standard FPT problems, one seeks to evaluate the FPT distribution $F(r,t)$, or its moments. The former is is the probability density that a particle, starting from the origin, will hit a specific target point located at the distance $r$ from the origin at time $t$. For our needs we generalize the problem, replacing the point-like target by an extended network-like target, as follows:

{\it Random walks are performed on a lattice $\mathcal L$ of dimension $d$ with targets forming a proper sublattice $\mathcal L_a\subset \mathcal L$ of dimension $d_a<d$. At $t=0$, an initial position is chosen randomly on $\mathcal L_a$. Find the distribution function $F(t)$ that a particle will return to $\mathcal L_a$ (not necessarily the initial position) for the first time at time $t>0$.}



Let $P_0(t)$ be a probability density for a particle to occupy the target sublattice $\mathcal L_a$ at time $t$, provided it was on $\mathcal L_a$ at $t=0$. 
(As in Section 2, the subscript $0$ indicates that $P_0(t)$ is evaluated with absorption turned off; the targets in the above FPT problem are not traps, perfect or otherwise.) Clearly, the FPT probability distribution $F(t)$ and the occupation probability distribution $P_0(t)$ are related in the same way as 
for a point-like target~\cite{Redner_book},
    \begin{eqnarray}
        P_0(t)=\delta(t)+\int_0^t F(\tau)\, P_0(t-\tau)\,d\tau.
        \label{FPT_eq}
    \end{eqnarray}
Here the first term on the right reflects the initial condition of being on $\mathcal L_a$ at $t=0$, and the second term says that to occupy $\mathcal L_a$ at time $t$ the particle must hit it for the first time at some moment $\tau<t$  and then return to $\mathcal L_a$ after time $t-\tau$. The corresponding equation for Laplace transforms (which we denote by tildes) reads $\tilde P_0(s)=1+\tilde F(s)\,\tilde P_0(s)$, so that
    \begin{eqnarray}
        \tilde F(s)=1-1/\tilde P_0(s).
        \label{F_laplace}
    \end{eqnarray}
As was shown in Section 2, the compact exploration argument
gives for $P_0(t)$ the asymptotic scaling (\ref{P0}),
$P_0(t)\sim t^{-1+\alpha}$. Then 
according to the Tauberian theorem~\cite{Feller},  
$\tilde P_0(s)\sim s^{-\alpha}$ for small $s$, and  from  (\ref{F_laplace}) one obtains $\tilde F(s)\sim 1-s^\alpha$. In the long time domain this corresponds to power law decay of the FPT distribution,
    \begin{eqnarray}
        F(t)\sim t^{-1-\alpha}.
        \label{FPT}
    \end{eqnarray}
For a single point-like target $d_a=0$, $\alpha=1-d/d_w=1-d_s/2$, and (\ref{FPT})  takes the form $F(t)\sim t^{-2+d_s/2}$. This specific form of the result (\ref{FPT}) was obtained and tested in~\cite{Meroz}.

\begin{figure}
  \includegraphics[height=6.2cm]{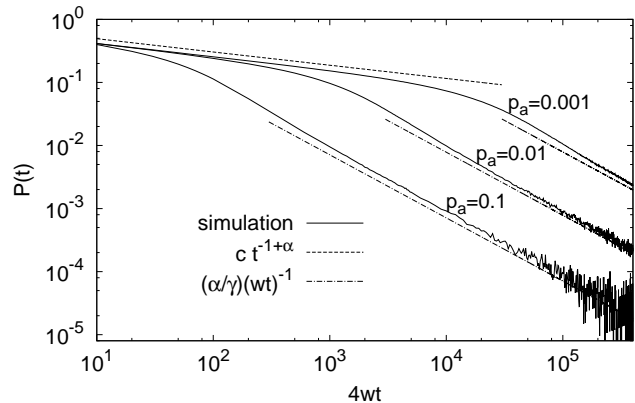}
  \caption{
  A log-log plot of the conditional probability $P(t)$ 
  of surviving until, and occupying a trap at, time $t$ (see Eq.~\ref{P_exact}).
  It is evaluated over $N\sim 10^8$ walks 
  on the 2D
  lattice with imperfect traps on the Sierpinski gasket $G_{20}$ (see Fig. 7) for several values
  of the absorption probability $p_a$. Solid lines show simulation results, the dashed line shows 
  the dependence (\ref{P0}), $P_0(t)=c\,(wt)^{-1+\alpha}$,
  and dash-dotted lines represent the ansatz $P(t)=\alpha\gamma^{-1}(wt)^{-1}$.
  }
  \label{fig_10}       
\end{figure}

With the FPT distribution found, we now return to the trapping problem. We identify the target network with the trap sublattice and appeal to the above reasoning that on long time-scales the time of absorption is approximately that of the time of first return on the trap sublattice, which is distributed according to (\ref{FPT}). The survival probability $f(t)$ is the probability that first passage occurs at $\tau>t$, so one immediately recovers the  power law (\ref{power}),
    \begin{eqnarray}
        f(t)\sim \int_t^\infty F(\tau)\,d\tau\sim t^{-\alpha}.
        \label{power2}
    \end{eqnarray}
Although it is not formally recovered in this derivation, the prefactor $1/p_a$ -- which we found experimentally; see (\ref{power}) -- is intuitively to be expected in (\ref{power2}).

One may alternatively seek to account for algebraic asymptotic decay $f(t) \sim t^{-\alpha}$ in terms of the conditional probability $P(t)$ of occupying a trap (\ref{P_exact}).
Namely, this behavior can be formally
derived as a solution of equation (\ref{f}), $df/dt=-w_a\,P(t)\,f$,
when $P(t)$ has the form
$P(t)=\alpha\,\gamma^{-1}(wt)^{-1}$.
Indeed, numerical simulation shows that at large time scales there is a crossover from
$P_0(t)\sim t^{-1+\alpha}$ to $P(t)\sim 1/t$; see Fig. 10. 
We find it difficult, however, to justify or interpret the above ansatz for $P(t)$ theoretically, and numerically
the function $P(t)$ is hard to evaluate; it  
quickly becomes very small and, unlike $f(t)$, fluctuates wildly even when evaluated over a very large number of walks.

\section{Conclusion}

Stretched exponential and power laws are the two most commonly observed
heavy-tailed distributions in  disordered and complex systems, and for this reason are often believed to originate from very general mechanisms. In particular, power law kinetics are often a signature of continuous-time random walk processes~\cite{MW} characterized 
by a broad distribution of
transition rates or waiting times, and there are several
models~\cite{Palmer,Huber,Sorn,Sturman,Phillips,Johnston,Campbell} still competing as generic explanations for the origin of stretched exponential kinetics. From this perspective, the emergence  of 
both these distributions within the same conceptually simple 
model studied in this paper is perhaps remarkable.
We studied the survival  probability $f(t)$ of random walks in the presence of fractally correlated traps.
For imperfect, weakly absorbing traps,
the initial relaxation of $f(t)$ is stretched exponential, followed by power law decay,
with both regimes characterized by the same exponent $\alpha$.
The regime of stretched exponential relaxation
is shorter for strongly absorbing traps, but may hold over several orders
of magnitude in time for weakly absorbing traps.
Both regimes may be accounted for by arguments based
on the concept of compact exploration, applied
to evaluate pertinent occupational and first-passage time distributions, for stretched exponential and power law regimes respectively.

We illustrated and verified theoretical predictions with
Monte Carlo simulations for regular host and trap lattices, but we
also expect these results for random fractal lattices like critical percolation
clusters~\cite{Havlin_book,Nakanishi} and multidimensional  potential landscape
structures relevant to complex systems with correlated
disorder~\cite{Stillinger}. In the latter case, imperfect correlated traps may
correspond to deep potential valley regions separated from the relaxation
pathway by a potential ridge. 

Theoretical arguments employed in the paper 
imply that the host and trap lattices are infinite, 
and in the simulation we tried to minimize the effects of 
boundary conditions.
The enumeration algorithms employed in this paper
allow one to simulate random walks on very large 
fractal structures. All results presented are for
fractals of generation $G\ge 20$, each consisting of at least $10^9$ units, which we found to be sufficiently large to neglect finite-size effects.
For the time scale considered ($z\,w\,t\alt 10^6$),
we found empirically 
that finite size effects become noticeable only for much smaller structures of generation $G<10$. 
While specific forms of finite size effects depend on boundary  conditions,  
we found as a general trend that they make the survival probability $f(t)$ decay faster at long times than  in an infinite system. 
This is intuitively clear, since 
in an infinite fractal system a particle finds itself, 
as time progresses, in larger and larger 
trap-free regions, whereas in a finite system the 
maximum size of a trap-free region is fixed.

 \begin{acknowledgments}
We thank G. Buck,  J. Schnick, S. Shea, and J. Parodi for discussions and interest, and the anonymous referees for their insightful comments and suggestions.
\end{acknowledgments}

\renewcommand{\theequation}{A\arabic{equation}}
\setcounter{equation}{0}

\section*{Appendix}

In this Appendix we outline a method for the evaluation of the survival
probability  $f(t)$  for a random walk on a one-dimensional lattice with an
imperfect trap located at the origin.
The problem is described by the master equation
(\ref{master2}). Using the standard Laplace and discrete Fourier
transforms, one can readily find from that  equation the exact expression for
the Laplace transform of $f(t)$, whose exact inverse however is unknown. More
analytic progress
can be achieved by considering, instead of the exact master
equation (\ref{master2}), its continuous limit version
    \begin{eqnarray}
        \frac{\partial }{\partial t}\,f(x,t)=D\,\frac{\partial^2}{\partial x^2}\,f(x,t)
        -r\, \delta(x)\, f(x,t),
        \label{A_master}
    \end{eqnarray}
for the probability density $f(x,t)$. This equation  follows from  (\ref{master2})
after replacements $x=n\Delta$,  $f(x,t)=f_n(t)/\Delta$,
$\delta(x)=\delta_{n0}/\Delta$,
and taking the limits $\Delta\to 0$ and $w,\,w_a\to\infty$ with finite 
    \begin{eqnarray}
        D=w\,\Delta^2,\qquad r=w_a\,\Delta.
        \label{A_D}
    \end{eqnarray}
 Let
    \begin{eqnarray}
        G(x,t)=\frac{1}{\sqrt{4\pi D t}}\,\exp\left(-\frac{x^2}{4Dt}\right)
        \label{A_propagator}
    \end{eqnarray}
be a free-diffusion propagator, that is, a solution of the trap-free  diffusion
equation (Eq.~(\ref{A_master}) with $r=0$) with the initial condition
$G(x,0)=\delta(x)$. Then the solution of Eq.~(\ref{A_master}) with initial
condition $f(x,t)=\delta(x-x_0)$ can be expressed as follows~\cite{Bru}:
    \begin{eqnarray}
        \!\!\!\!\!
        f(x,t)=G(x-x_0,t)-r\!\int_0^t \!f(0,t')\,G(x,t-t')\,dt'.
        \label{A_sol1}
    \end{eqnarray}
This expression is easy to interpret: $f(x,t)$ is smaller  than the
free-diffusion propagator $G(x-x_0,t)$ by the contribution from the particles,
which  were captured by the trap at an earlier time $t'<t$ and, had they not been captured, would have diffused to the point $x$ at time $t$. The negative
contribution of such particles is given by the second
term in the right-hand-side of (\ref{A_sol1}).

In the  Laplace space, Eq.~(\ref{A_sol1}) reads
    \begin{eqnarray}
        \tilde f(x,s)=\tilde G(x-x_0,s)-r\,\tilde f(0,s) \,\tilde G(x,s),
        \label{A_sol2}
    \end{eqnarray}
where the tilde denotes Laplace transforms. From here one finds
    \begin{eqnarray}
        \tilde f(0,s)=\frac{\tilde G(x_0,s)}{1+r\,\tilde G(0,s)},
        \label{A_sol3}
    \end{eqnarray}
and substituting this expression back to (\ref{A_sol2}) one gets
    \begin{eqnarray}
        \tilde f(x,s)=\tilde G(x-x_0,s)-\frac{r\,\tilde G(x_0,s)}{1+r \,\tilde G(0,s)}\,\tilde G(x,s).
        \label{A_sol4}
    \end{eqnarray}
For the Laplace transform of the survival probability this yields
    \begin{eqnarray}
        \tilde f(s)=\int_{-\infty}^\infty\tilde f(x,s)\,dx=\frac{1}{s}\left(
        1-\frac{r\,\tilde G(x_0,s)}{1+r\,\tilde G(0,s)}
        \right).
    \end{eqnarray}
Substituting the Laplace transform of the propagator (\ref{A_propagator})
    \begin{eqnarray}
        \tilde G(x,s)=\frac{1}{\sqrt{4Ds}}\,\exp\left(-|x|\,\sqrt{\frac{s}{D}}\right)
    \end{eqnarray}
one finds
    \begin{eqnarray}
        \tilde f(s)=\frac{1}{s}\left[
        1-\frac{\exp\left(-|x_0|\sqrt{\frac{s}{D}}\right)}
        {1+\frac{1}{r}\sqrt{4Ds}}
        \right].
\label{A_image0}
    \end{eqnarray}
One  expects this result, obtained in the continuous limit, to
be a reasonable approximation for a discrete lattice as well.  
For that case, taking into account (\ref{A_D}),
we can re-write (\ref{A_image0}) using notation from the main text, 
    \begin{eqnarray}
        \tilde f(s)=\frac{1}{s}\left[
        1-\frac{\exp\left(-i_0\,\sqrt{\frac{s}{w}}\right)}
        {1+\frac{2}{\gamma}\sqrt{\frac{s}{w}}}
        \right],
        \label{A_image}
    \end{eqnarray}
where $i_0=|x_0|/\Delta$ is the initial position in lattice spacing units,
and $\gamma=w_a/w$ is the dimensionless parameter characterizing the absorption strength.
While this Laplace transform enjoys an exact closed-form inverse,
see Eq. (\ref{A_exact}) below, the asymptotic forms of $f(t)$ can be derived from that of   (\ref{A_image}).

Consider first the case of weak absorption, 
\begin{eqnarray}
\gamma\ll 1, \qquad i_0\ll 1/\gamma.
\end{eqnarray}
Then for the domain
\begin{eqnarray}
i_0\ll \sqrt{w/s}\ll 1/\gamma
\label{A_domain1}
\end{eqnarray}
(\ref{A_image}) can be approximated as
    \begin{eqnarray}
        \tilde f(s)\approx s^{-1}
        -\frac{\gamma}{2}\,w^{1/2}\, s^{-3/2}.
        \label{A_approx1}
    \end{eqnarray}
Interval (\ref{A_domain1}) corresponds to the time domain
\begin{eqnarray}
i_0^2\ll wt\ll \gamma^{-2},
\label{A_domain2}
\end{eqnarray}
for which the inversion of (\ref{A_approx1}) gives 
    \begin{eqnarray}
        f(t)\approx 1-\frac{1}{\sqrt{\pi}}\,\gamma\, \sqrt{w\,t}.
        \label{A_analytic}
    \end{eqnarray}
This is a good short-time approximation of the stretched exponential function (\ref{ser2}). 
On the other hand, for $\sqrt{w/s}\gg 1/\gamma$ (\ref{A_image}) is 
reduced to 
\begin{eqnarray}
\tilde f(s)\approx \frac{2}{\gamma}\,\frac{1}{\sqrt{w\, s}}.
\label{A_qq}
\end{eqnarray}
In the time domain this corresponds to a power law asymptotics 
\begin{eqnarray}
f(t)\approx \frac{2}{\sqrt{\pi}}\,\frac{1}{\gamma}\, \frac{1}{\sqrt{wt}}.
\label{A_analytic2}
\end{eqnarray}
for $wt\gg\gamma^{-2}$.

If absorption is not small, $\gamma\agt 1$,  
the conditions (\ref{A_domain2}) and $i_0\ll 1/\gamma$ 
become inconsistent, 
and the stretched exponential regime is absent. In this case,
for 
\begin{eqnarray}
\sqrt{w/s}\gg 1/\gamma, i_0
\end{eqnarray}
one obtains from (\ref{A_exact}) instead of (\ref{A_qq}),
the approximation
\begin{eqnarray}
\tilde f(s)\approx \left(i_0+\frac{2}{\gamma}\right)\,\frac{1}{\sqrt{w\, s}}.
\label{A_qqq}
\end{eqnarray}
In the time domain this corresponds to 
\begin{eqnarray}
f(t)\approx \frac{1}{\sqrt\pi}\,\left(i_0+\frac{2}{\gamma}\right)\,\frac{1}{\sqrt{w\, t}}.
\label{A_analytic3}
\end{eqnarray}
for $wt\gg i_0, 1/\gamma$. In particular, for the limit of a perfect
trap $\gamma\to\infty$, 
\begin{eqnarray}
f(t)\approx \frac{i_0}{\sqrt\pi}\,\frac{1}{\sqrt{w\, t}}, \qquad wt\gg i_0.
\label{A_analytic4}
\end{eqnarray}

The exact inverse of (\ref{A_image}) has the form~\cite{Kenkre}:
\begin{eqnarray}
f(t)&=&
\exp\left\{\,
\frac{\gamma}{2}\,i_0+\frac{\gamma^2}{4}\,wt\,
\right\} \erfc\left\{\,
\frac{1}{2}\,\frac{i_0}{\sqrt{wt}}+\frac{\gamma}{2}\,\sqrt{wt}\,
\right\}\nonumber\\
&+&\erf\left\{
\frac{1}{2}\,\frac{i_0}{\sqrt{wt}}
\right\}.
\label{A_exact}
\end{eqnarray}
(In the limit of a perfect trap $\gamma\to\infty$ only the
second term survives in this expression.)
The above asymptotic formula 
may alternatively be obtained directly from (\ref{A_exact}) 
using asymptotic relations
$e^{x^2}\erf(x)\approx (2/\sqrt{\pi})\,x$ for $x\ll 1$ and
$e^{x^2}\erfc(x)\approx1/(\sqrt{\pi}\,x)$ for $x\gg 1$.

Another way to treat the problem is to exploit, instead of
Eq.(\ref{A_master}), the trap-free diffusion equation with
the radiation boundary condition
$\partial f(x,t)/\partial x|_{x=0}= k\,f(x,t)|_{x=0}$ with $k=r/(2D)$~\cite{Bru,Tait,BNaim}.

\end{document}